\documentclass[letterpaper,prl,twocolumn,showpacs]{revtex4}
\usepackage{graphicx}
\begin{document}

\title{Experimental demonstration of a squeezing enhanced power recycled\\
Michelson interferometer for gravitational wave detection}
    
\author{Kirk McKenzie} 
\affiliation{Gravitational Wave
Research Facility, Department of Physics, Faculty of Science,
Australian National University, ACT 0200, Australia} 
\author{Daniel A. Shaddock} 
\affiliation{Gravitational Wave Research Facility,
Department of Physics, Faculty of Science, Australian National
University, ACT 0200, Australia} 
\author{David E. McClelland} 
\affiliation{Gravitational
Wave Research Facility, Department of Physics, Faculty of Science,
Australian National University, ACT 0200, Australia}
\author{Ben C. Buchler}
\affiliation{Quantum Optics Group, Department of Physics, Faculty of
Science, Australian National University, ACT 0200, Australia}
\author{Ping Koy Lam} 
\affiliation{Quantum Optics Group, Department of
Physics, Faculty of Science, Australian National University, ACT 0200,
Australia} 

\begin{abstract}
Interferometric gravitational wave detectors are expected to be limited by shot noise at some frequencies.  We experimentally demonstrate that a power recycled Michelson with squeezed light injected into the dark port can
overcome this limit.  An improvement in the signal-to-noise ratio of 2.3dB is measured and locked stably for long periods of time.  The configuration, control and signal readout of our experiment are compatible with current gravitational wave detector designs. We consider the application of our system to long baseline interferometer designs such as LIGO.
\end{abstract}

\pacs{04.80.Nn, 07.60.Ly, 95.55.Ym}
\date{\today} \maketitle

The first generation of interferometric gravitational wave (GW)
detectors are expected to begin taking data in 2002.  Although they
will be the most sensitive devices ever built, they are predicted to detect only large, infrequent gravitational events.  To regularly detect sources, and thereby allow comparison with astrophysical models, a factor of ten improvement in sensitivity is required.  The second generation of detectors, such as Advanced LIGO \cite{whitepaper}, is expected to reach this goal.   These instruments will
employ 100~W class lasers, light recycling techniques, fused silica suspension systems and high mechanical
$Q$ sapphire mirror substrates.  Early predictions are that they will
be limited by quantum noise, i.e. noise arising from the quantum
nature of light, over most of the GW signal frequency band (10Hz to
1000Hz) \cite{whitepaper}.  This has sparked an explosion of theoretical
proposals on the application of quantum optical techniques to surpass
the quantum limit in laser interferometry. These include:
the use of squeezed states \cite{Kimble2} building on the proposal by Caves \cite{Caves}; quantum nondemolition schemes
including changing of the readout parameters \cite{Braginsky} and
 opto-mechanical coupling techniques \cite{Buononno&Chen}. 
What makes these theoretical proposals even more exciting is the maturity of experimental techniques for generating non-classical states of light.  In 1985, the landmark experiment of Slusher \emph{et al.}\cite{Slusher} measured 0.3dB of quantum noise suppression. Bench-top experiments currently produce over 7dB reduction of the quantum noise \cite{Schneider98.OE,Lam99.JOB}. A combination of current squeezing
technology with the high power and high stability of GW detection
laser and optical systems now makes 10dB of squeezing a realistic goal
\cite{Kimble2}.

Despite the potential for squeezing to improve interferometric
sensitivity, to date there has been no experimental demonstration of
squeezing applied to an interferometer bearing any resemblance to a GW
detector.  Squeezing enhanced performance has been demonstrated in
other interferometers, such as the Mach-Zehnder \cite{Kimble} and
polarimeter \cite{Grangier}.  None of these experiments employed a
Michelson configuration; used light recycling techniques; or utilised
a signal readout scheme compatible with an advanced GW detector. 
Theoretical analysis of Gea-Banacloche {\it et al.} \cite{Leuchs} suggested
that squeezing is broadly compatible with recycling techniques. 
However, the difficulty in devising a readout and control scheme
compatible with both squeezing and light recycling has, until now,
prevented any definitive demonstration.

In this letter, we report an experimental demonstration of a power
recycled Michelson \cite{PRM2} with locked optical squeezing injected into the dark port. Our source of squeezing was an optical parametric amplifier (OPA). We used a laser system, configuration, control and
readout system compatible with advanced GW detector proposals.  The entire system maintained lock for long periods and we measured a signal
with noise below the shot noise limit (SNL).   Contrary to previous work, we found
that there is a link between power recycling and squeezing.  In fact,
the efficiency of squeezing detected at the dark port with power
recycling can be higher than for a simple Michelson.  Given the
success of our experiment, we review theoretically the compatibility
and application of our system to long base-line 
detectors, such as LIGO \cite{LIGO}.

\begin{figure}[t!]
  \begin{center}
  \includegraphics[width=8cm]{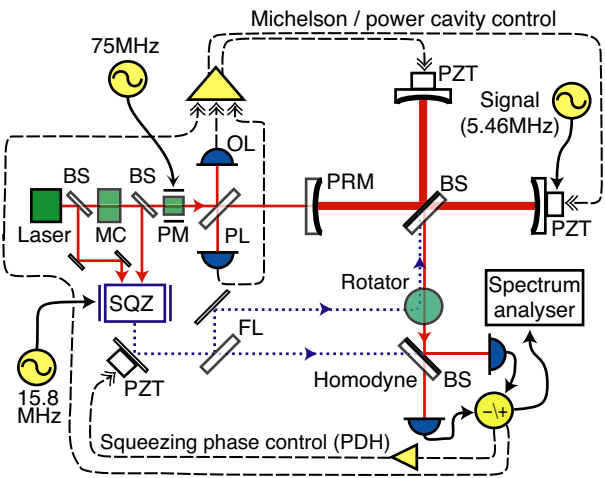}
  \end{center}
  \caption{Schematic of the experiment.  Squeezed light can be
  injected into the output port of the interferometer via the Faraday
  rotator or interrogated using homodyne detection.  The control servos for the interferometer are shown schematically by the dashed lines.  BS=beam-splitter, CL= power cavity locking detector FL=flipper mirror, OL=offset locking detector, PM=phase modulator (75MHz),  PZT=piezo-electric transducer and SQZ=squeezed state generation system.}
  \label{setup}
\end{figure}

The experimental setup is shown in Fig.\ref{setup}. Of the 700mW of power generated by our laser (Nd:YAG 1064nm), about 70\% was used to pump the squeezing system. The remainder passed through a mode-cleaning cavity \cite{Rudiger}, giving an output beam which was shot-noise limited above 4MHz.  This beam provided both the interferometer input and a seed wave for the OPA. The low power ($\sim10\mu$W) squeezed output of the OPA was phase modulated at 15.8MHz.  Details of the OPA system may be found in Ref. \cite{Buchler}. The squeezed light was injected into the interferometer via a Faraday rotator (Gs\"{a}nger FR 1060/5) to give spatial degeneracy between the injected squeezing and the output beam of the interferometer, as required in a long baseline GW detector.  The fringe visibility between the squeezing and the interferometer input beam was 99\%.  The squeezed beam could also be measured directly by the homodyne system via a flipper mirror.  The interferometer had a power cavity of length 1m and power mirror of  90\% reflectivity.  The input beam to the interferometer had a power of 20mW, of which 3mW exited at the dark port where it was measured by the homodyne system. The homodyne detectors were built around ETX 500 photodiodes with 93\% quantum efficiency. A GW signal was simulated by modulating one of the Michelson arm lengths using a piezo-electric transducer (PZT) at a frequency outside the power cavity bandwidth. In our case this was 5.46MHz \cite{note}.  The signal readout was obtained from the sum output of the homodyne detectors.

A vital feature of our setup is that the control scheme is
compatible with future GW detectors and squeezed light.  For squeezing to be of any use, the squeezed quadrature must be in phase with the interferometer
output.  The 15.8MHz phase modulation of the squeezed beam allowed us to use the Pound-Drever-Hall (PDH) \cite{PDH} method to gather an error signal for the squeezed beam phase.  This was fed-back to a PZT in the squeezed beam path. The power cavity error signal was also derived using the PDH technique (using the 75MHz phase modulation) and was fed-back to the length of the power cavity, via PZTs on the Michelson mirrors.   The relative length of the Michelson arms was controlled with offset locking. 
This error signal was derived by subtracting an offset voltage, obtained via the detector OL, from the sum voltage of the homodyne detectors \cite{offset}. 
\begin{figure}[t!]
  \begin{center}
  \includegraphics[width=8cm]{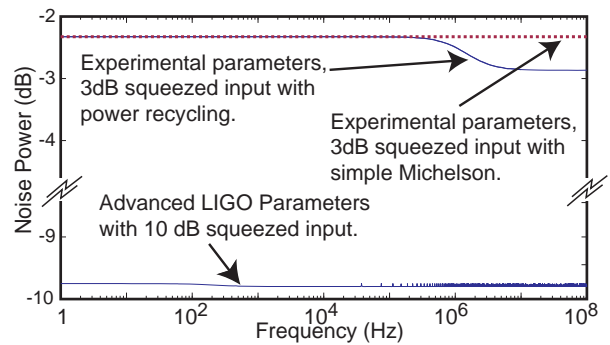}
  \end{center}
  \caption{Predicted shot noise suppression at the dark port is shown as a function of frequency.  The vertical scale is dB relative to the SNL (at 0dB). 
Top traces:  a model of our experiment assuming 3dB of squeezing, power mirror reflectivity of 90\% and 15\% rotator loss. Bottom trace: Predicted performance with Advanced LIGO parameters, 10dB of squeezing and 5\% rotator loss.} \label{theory}
\end{figure}
The error signal was fed-back to PZTs on the Michelson mirrors.  Offset locking is being used in Advanced LIGO prototypes
\cite{strain} and is highly compatible with
squeezing.  It requires no modulation, so that unlike Schnupp/frontal
modulation \cite{Schnupp}, the signal readout does not require
demodulation.  This is a benefit as the squeezing is only required at
the GW signal frequencies.  If modulation techniques are used,
squeezing is required at the signal frequency and twice the
modulation frequency $\pm$ the signal frequency \cite{chickarmane}. 
This simplification of the requirements for squeezing in a GW detector
is valuable.

Squeezed states of light may also be used to suppress radiation
pressure noise in GW detectors by tuning the angle of the squeezed
quadrature \cite{Pace93.PRA,Kimble2}.  Our experimental setup can, in
principle, be used to suppress radiation pressure noise, however our
experiment does not operate in a regime where radiation pressure is
significant.  Suppression of radiation
pressure noise is, therefore, not considered in this letter.

Shot noise arises as the electromagnetic vacuum mode enters the dark
port of the beam-splitter.  The vacuum mode interferes with the input
beam passing on its noise characteristics.  The smallest phase a SNL
interferometer can measure is:
\begin{equation}
    \Delta \phi \ge 1/\sqrt{n},
\end{equation}
where $n$ is the number of photons incident on the beamsplitter. 
Sub-shot noise sensitivity can be achieved if the vacuum mode is
filled by a squeezed state.  The noise suppression can be
characterised by the intensity variance of the light at the
interferometer output, $V_{\rm pd}$ (where the variance is normalised
to the SNL).
When optimised, this variance is a function of the input beam
intensity noise, $V_{\rm LO}$;  the squeezed state noise, $V_{\rm
sqz}$ and the losses in the interferometer, modelled by adding vacuum
noise with variance $V_{v}=1$.  The normalised transfer functions for
each noise input to the photodetector for our system are $T_{\rm
LO}(\omega), T_{\rm sqz}(\omega)$ and $T_{v}(\omega)$, so that the
detected variance $V_{\rm pd}$ is given by
\begin{equation}    
V_{\rm pd}=|{T_{\rm LO}(\omega)}|^{2}V_{\rm LO}+|{T_{\rm 
sqz}(\omega)}|^{2}V_{\rm sqz} +|{T_{v}(\omega)}|^{2}V_{v}.
\label{trans}
\end{equation}
\begin{figure}[t!]
  \begin{center}
  \includegraphics[width=8cm]{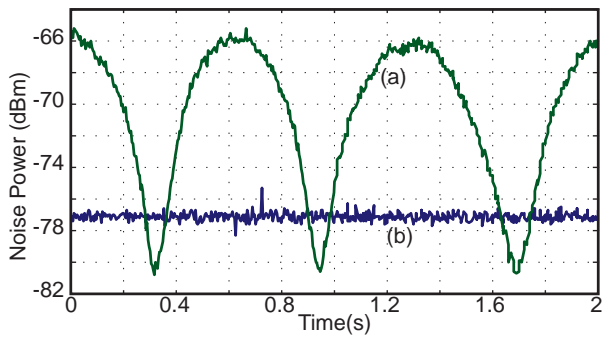}
  \end{center}
  \caption{a) Scan of squeezing at 5.5~MHz measured before the interferometer by homodyning; b) Shot noise at 5.5~MHz. Electronic noise lies at -84.9dBm, Resolution bandwidth (RBW) = 100kHz, Video Bandwidth (VBW) = 300Hz}
  \label{scan}
\end{figure} 
These transfer functions are frequency independent for a simple
Michelson.  If a power recycling mirror is present, they are functions of the
power recycling cavity linewidth.  When the interferometer is locked
close to a dark fringe, the intensity fluctuations on the input beam
couple weakly to the interferometer output, i.e. $T_{\rm LO}$ is very small. 
The amount of squeezing that couples to the output, however, is
large as ${T_{\rm sqz}(\omega)}$ is close to 1.  This is the dominant
term in equation (\ref{trans}).  To model our experiment we consider
3dB of input squeezing over the detection frequency range. 
Theoretical plots of the frequency spectra of the detected output
variance for a simple Michelson and a power recycled Michelson (PRM)
are shown in Fig.\ref{theory}.  The results show improved
performance of the PRM compared to the simple Michelson.  This
difference is a result of two effects.  Firstly, to keep the
same power at the homodyne photodetector, the position of the Michelson fringe shifts, such that the effective Michelson reflectivity increases.  In our experiment the reflectivity for the simple Michelson is $\approx 0.92$, where as it is $\approx 0.99$ for the PRM. The result
is an increase in $T_{\rm sqz}(\omega)$ and a decrease in $T_{\rm
LO}(\omega)$, so that more squeezing is transferred to the interferometer output.  In the presence of a power recycling mirror, therefore,
less squeezing is wasted in the interferometer.  Secondly, the power
mirror introduces a frequency dependence to the squeezing transfer function, $T_{\rm sqz}$. Outside the power cavity linewidth, the interferometer becomes highly reflective so that transfer of the squeezing becomes close to ideal.

Figure \ref{scan} shows the noise of our squeezed beam prior to entering the interferometer. The noise suppression is $\sim$3.5dB below the SNL.  When the squeezing was injected into the interferometer, our
Faraday rotator gave 15\%  loss (double pass).  This was the
dominant source of loss in our experiment.  It reduced the amount of squeezing coupled into the interferometer to $\sim$2.8~dB (measured).

\begin{figure}[t!]
  \begin{center}
  \includegraphics[width=8cm]{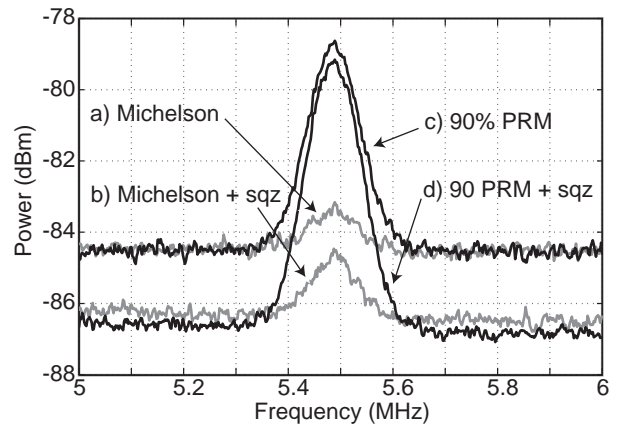}
  \end{center}
  \caption{Results showing sub-shot noise sensitivity in our interferometer, with and without power recycling. Electronic noise lies at -93.5dBm, The signal is at 5.46~MHz .  RBW = 100 kHz, VBW = 30 Hz}
  \label{90prm}
\end{figure}

A signal at 5.46~MHz was used to characterise our Michelson
interferometer.  Fig.~\ref{90prm} trace (a) shows the result for a simple
Michelson.  The noise floor is at the SNL. When squeezing was
introduced (trace (b)) we observed noise suppression of 1.8 $\pm $0.2~dB below
the SNL. Traces (c) and (d) show the response of the PRM with and without
the squeezed input.  The signal power is proportional to the circulating power.  Our power recycling factor of $\approx$~4 therefore gives the PRM a signal $\approx$~4 times larger than the simple Michelson.  The noise floor of trace (d) is 2.3 $\pm$ 0.2~dB lower than SNL.  The predicted improvement of the squeezing performance with power recycling is evident in the different noise floors of traces (b) and (d).  If the electronic noise is subtracted from trace (d), the squeezed noise floor is found to lie 3.0 $\pm$ 0.2~dB below SNL. Our experiment maintained lock for periods longer than 15 minutes, and was limited only by the temperature stability of our laboratory. 

These results confirm that squeezing is compatible with power recycling, however the practicality of using squeezed light in existing and future GW detectors depends on the compatibility with current designs and the amount of squeezing that will be actually available.  We will briefly look at these two issues.

In terms of compatibility to proposed GW detectors, there are three
outstanding considerations: the requirement of arm cavities; signal recycling
and the generation of squeezing in the GW signal band. The arm cavities
introduce fractionally higher loss into the interferometer ($\sim$1\%
\cite{Kimble2}), and require additional locking servos.  The extra
loss will result in only $\sim$0.1dB reduction of the squeezing, while the
locking of the arm cavities will be independent of the squeezing.

With regard to signal recycling, modelling of a system with long
signal storage times \cite{chickarmane} has shown that the squeezing
will sense an impedance matched cavity and therefore be transmitted
through the interferometer at the signal frequency.  Squeezing is
nonetheless advantageous as it will broaden the bandwidth of the
signal recycling.  With resonant sideband extraction (RSE) \cite{RSE},
the injected squeezing will sense a reflective cavity.  In this case
we do not expect significant degradation of squeezing due to signal recycling. 
Employing squeezing and signal recycling together, while feasible,
might not be a worthwhile exercise.  Squeezing is only helpful if quantum noise sources are limiting the interferometer performance. A combination of RSE and squeezing is likely to reach a regime where other noise sources, particularly thermal noise, will be dominant at low frequencies \cite{whitepaper}.  In this case, squeezing will only be of benefit at higher frequencies, outside the thermal noise bandwidth.  Instead, one may imagine a squeezed, power recycled interferometer with arm cavities as an alternative to RSE \cite{Kimble2,McClellandTAMA}.

With respect to the squeezing bandwidth, our experiment shows squeezing
at MHz, rather than the $\sim$100Hz range required for GW detection. 
The major impediment to bench-top low frequency squeezing is laser
technical noise.  The lowest squeezing frequency in a continuous wave beam
reported to date is 200kHz, which was achieved by the use of two
 OPAs, enabling common-mode rejection of the classical noise
\cite{Bowen02.OL}.  Fortunately, GW detection facilities provide the
ideal environment for low frequency squeezed state generation.  In
particular there will be a ready supply of laser light shot noise 
limited to frequencies as low as 10Hz \cite{strain}, suitable for 
generation of squeezing inside the GW detection bandwidth.

The key issue governing the amount of squeezing that can be realistically injected into a GW detector is the availability of low loss optics, both for squeezing generation and application to the interferometer.  The dominant loss in our
interferometer was the 15\% double pass transmission of the
rotator.  Given the low absorption of the materials used in
rotators (such as TGG with absorption of  0.55\% cm$^{-1}$), careful
tuning of the magnetic fields around the crystal should provide
rotators with $<$ 5\% double pass loss.  The amount of squeezing generated is limited by material losses in the OPA crystal.  The high
laser powers available in GW facilities will allow squeezing
generation in regimes not possible in bench-top system.  High laser
powers may be used to pump OPA cavities with much lower finesse, 
thereby reducing the loss of the squeezing that occurs inside the OPA cavity. Given that bench-top squeezing experiments currently measure 7~dB of quantum noise suppression \cite{Schneider98.OE,Lam99.JOB}, over 10dB of squeezing is a realistic target in a GW facility.

In summary, the technology of high precision interferometry and
squeezed state generation, have now reached a stage where their fusion
presents a genuine alternative configuration for future gravitational
wave detectors.  Our results, which show the compatibility of advanced
interferometer design and squeezed light, represent the first step
toward this goal.

The authors would like to acknowledge Malcolm Gray for the design of the electronics used in our experiment and the funding of the Australian Research Council.


\begin{thebibliography}{99}
\bibitem{note} A bench-top interferometer is limited to MHz
signals due to the short power cavity.  The power cavity linewidth of
a long base-line interferometer will be of order 100~Hz.
\end{thebibliography}
\end{document}